\documentclass[aps,preprint,prd,showkeys,showpacs,nofootinbib]{revtex4}

\usepackage{latexsym}
\usepackage{amsmath}
\usepackage{graphicx}
\usepackage{subfigure}
\usepackage{dcolumn}
\usepackage{bm}
\usepackage{amssymb}
\usepackage{latexsym}
\usepackage{ulem}
\usepackage{datetime}
\usepackage{scrtime}
\usepackage{color}
\usepackage[dvipsnames, svgnames, x11names]{xcolor}
\usepackage[colorlinks,linkcolor=blue, anchorcolor=blue, citecolor=Blue]{hyperref}

\def\be{\begin{equation}}
\def\ee{\end{equation}}
\def\ba{\begin{eqnarray}}
\def\ea{\end{eqnarray}}

\def\blue{\color{blue}}

\def\nn{\nonumber}
\def\lf{\left}
\def\rt{\right}

\begin{document}


\title{ Pre-inflation and Trans-Planckian Censorship}

\author{Yong Cai$^{1}$\footnote{\texttt{\blue caiyong13@mails.ucas.ac.cn}}}
\author{Yun-Song Piao$^{1,2}$\footnote{\texttt{\blue yspiao@ucas.ac.cn}}}

\affiliation{$^1$ School of Physics, University of Chinese Academy of
Sciences, Beijing 100049, China}

\affiliation{$^2$ Institute of Theoretical Physics,
Chinese Academy of Sciences, Beijing 100190, China}

\begin{abstract}

We investigate the implication of Trans-Planckian Censorship Conjecture (TCC) for the initial state of primordial perturbations. It is possible to set the state of perturbation modes in the infinite past as the Minkowski vacuum, only if the pre-inflationary era is past-complete. We calculate the evolution of the perturbation modes in such a pre-inflationary era and show that at the beginning of inflation the perturbation modes with wavelengths much shorter than the Hubble scale (but still larger than the Planck length scale) will behave as they are in the Bunch-Davis state. Therefore, a past-complete pre-inflationary evolution may automatically prepare the initial state required for the inflationary perturbations at the CMB window while obeying the TCC.

\end{abstract}



\maketitle

\tableofcontents

\section{Introduction}

Inflation
\cite{Guth:1980zm,Linde:1981mu,Albrecht:1982wi,Starobinsky:1980te}
is the standard paradigm of the early universe. It predicted the
nearly Gaussian and scale-invariant primordial perturbations, which
are consistent with the observations of cosmic microwave background
(CMB) anisotropies \cite{Ade:2015xua,Ade:2015lrj}. However,
inflation is not the final story of the early universe, since it
is past-incomplete, as shown in
Refs. \cite{Borde:1993xh,Borde:2001nh}.

Usually, the initial state of the inflationary perturbations is
set as the Bunch-Davis (BD) vacuum \cite{Bunch:1978yq}, which is
the lowest-energy state of de Sitter spacetime. However, due to
the past-incompleteness of inflation \cite{Borde:2001nh},
initially, the background and fluctuations of spacetime will be
inevitably involved in the cosmological singularity, where the
length scale is smaller than the Planck scale or the cutoff scale
of the effective field theory (EFT). This is the so-called
``trans-Planckian" problem of inflation
\cite{Brandenberger:2000wr,Martin:2000xs}. In this situation,
inflation does not automatically pick out the BD state as the
initial state of perturbation modes.\footnote{Non-BD initial
states may significantly affect the power spectrum and
Non-Gaussianity of primordial perturbations, e.g.,
\cite{Ashoorioon:2004vm,Ashoorioon:2005ep,Chen:2013tna,Ashoorioon:2013eia,Ashoorioon:2014nta,Ashoorioon:2018sqb}.}

Recently, Bedroya and Vafa proposed the Trans-Planckian
Censorship Conjecture (TCC) \cite{Bedroya:2019snp}, which states
that ``\textit{a field theory consistent with a quantum theory of
gravity does not lead to a cosmological expansion where any
perturbation with length scale greater than the Hubble radius
trace back to trans-Planckian scales at an earlier time}", see
also its application to inflation models \cite{Bedroya:2019tba}.
Actually, since inflation is past-incomplete, the
inflationary perturbations inevitably originated from the
sub-Planckian scale fluctuations. It is well-known that the BD
initial state of perturbations is quite essential for acquiring
the nearly scale-invariant power spectrum of density
perturbations. Therefore, how to set the BD state at the beginning
of the observable inflationary era (i.e., the CMB window) is a
question to be answered, see also \cite{DiTucci:2019xcr}. In a
certain sense, the TCC indicates that the existence of a
pre-inflationary era might be required.

One possibility of such pre-inflationary scenarios is a
nonsingular bounce, i.e., the so-called bounce inflation
\cite{Piao:2003zm,Liu:2013kea,Qiu:2015nha}, which not only avoids
the past-incompleteness problem of inflation but also reserves the
advantages of inflation. Recently, the instability problems of
perturbations in flat nonsingular cosmologies (which is inevitable
\cite{Libanov:2016kfc,Kobayashi:2016xpl} in Horndeski theory, see
also \cite{Ijjas:2016vtq,Dobre:2017pnt}) have been solved in the
``beyond Horndeski" theory, as proved within the EFT of
cosmological perturbations
\cite{Cai:2016thi,Creminelli:2016zwa,Cai:2017tku,Cai:2017dyi,Kolevatov:2017voe}, see also
\cite{Ijjas:2016tpn,Kolevatov:2016ppi,Mironov:2018oec,Boruah:2018pvq,Ye:2019frg,Ye:2019sth,Mironov:2019haz,Akama:2019qeh}
(see \cite{Langlois:2018dxi,Kobayashi:2019hrl} for reviews).\footnote{The earlier attempts (but with the instabilities
of perturbations) on nonsingular bounce, as well as G-bounce, can
be found in, e.g., Refs.
\cite{Cai:2007qw,Cai:2008qw,Creminelli:2006xe,Qiu:2011cy,Easson:2011zy,Cai:2012va,Cai:2013kja,Qiu:2013eoa,Koehn:2013upa,Battarra:2014tga,Koehn:2015vvy,Wan:2015hya,Odintsov:2014gea,Banerjee:2016hom}.}

In the pre-inflationary era, initially the universe is slowly
contracting with $\epsilon=-{\dot H}/H^2\gg 1$. Thus, in the
infinite past, the spacetime is nearly Minkowskian.\footnote{See also \cite{Liu:2014tda,Pirtskhalava:2014esa,Kobayashi:2015gga}
for the scenario with a slow expansion ($\epsilon\ll-1$) preceding
inflation.} In this case, it is natural to pick out the Minkowski
vacuum state as the initial state of the primordial perturbations.
However, at present, the pre-inflationary physics is still in
speculation and exploration. It is possible that the
pre-inflationary era might consist of multiple different phases.
The question is: provided the primordial perturbation modes
originate from the Minkowski vacuum  in the infinite past
and obey the TCC throughout, can the sub-horizon modes at the
beginning of inflation be in the BD state?
In this paper, we will investigate the relevant issues.

The outline of this paper is as follows:
In Sec.
\ref{Sec:T}, we briefly revisit the ``trans-Planckian" problem
within the EFT of cosmological perturbations. In Sec.
\ref{sec:ini}, we discuss the TCC and its implication. We point
out that it is possible to set the state of perturbations in the
infinite past as the Minkowski vacuum only if the spacetime is
past-complete. In Sec. \ref{sec-pert}, we calculate the evolution
of the perturbation modes in the past-complete pre-inflationary era to
see whether the corresponding modes approximate to the BD state at
the beginning of inflation. Sec. \ref{conclusion} is the
conclusion.

\section{ Trans-Planckian problem}
\label{Sec:T}

In this section, we will focus on the evolution of the primordial perturbations. In unitary gauge, the
quadratic action of scalar perturbation $\zeta$ is \be
S^{(2)}_\zeta =\int \, a^{3} Q_s\left[ \dot{\zeta}^{2} -c_s^2
\frac{\left(\partial_{i} \zeta\right)^{2}}{a^{2}}\right]d^{3} x d t\,.\label{eq:quadratic} \ee It is convenient to define
$u_k=z\zeta_k$ with $z=a \sqrt{2 Q_{s}}$, where $\zeta_k$ is the
scalar perturbation mode. The Mukhanov-Sasaki equation for $u_k$
is \be u_k^{\prime \prime}+\left(c_{s}^{2} k^{2}-\frac{z^{\prime
        \prime}}{z}\right) u_k=0\,,\label{MS-eq02} \ee where $'=d/d\tau$,
$c_s$ is the sound speed. In general relativity (GR), $c_s^2=1$ and $Q_s=
M_{\mathrm{P}}^2\epsilon$ with $\epsilon=-{\dot H}/H^2$.

In the slow-roll inflation, ${z^{\prime \prime}}/{z}\sim
a^{\prime\prime}/a\simeq (2+{\cal O}(\epsilon))/\tau^2$.
Thus the solution of Eq. (\ref{MS-eq02}) is \be
u_{k}(\tau)={\sqrt{-\pi \tau}\over 2}\lf[\alpha(k)
H_{\nu}^{(1)}(-k\tau)+\beta(k) H_{\nu}^{(2)}(-k\tau) \rt]\,, \ee
where the $k$-dependent coefficients $\alpha(k)$ and $\beta(k)$
obey the Wronskian condition $|\alpha|^2-|\beta|^2=1$,
$\nu\approx 3/2$, $H_\nu^{(1)}$ and $H_\nu^{(2)}$ are the first and
second kind Hankel functions of the $\nu-$th order, respectively.

One must select the initial state of perturbations to calculate the
coefficients $\alpha(k)$ and $\beta(k)$. Requiring the
perturbation modes coincide with the Minkowski solution ($u_k\simeq
{1\over \sqrt{2k}}e^{-ik\tau}$) in the infinite past
gives\footnote{Note that there could be an undetermined phase
    difference in $\alpha$ (see, e.g., \cite{Wang:2013eqj} for a
    review).} \be |\alpha|=1\,,\qquad |\beta|=0\,, \ee
where we have used that
\be 
H^{(1)}_\nu(-k\tau)\approx \sqrt{2\over -\pi
        k\tau}e^{-ik\tau} e^{-i\lf({\nu\over2}+{1\over4}
        \rt)\pi} 
\ee
in the limit $-k\tau\gg1$.
Particularly, we have $\nu=3/2$ for de Sitter inflation. Using $H^{(1)}_{3/2}(-k\tau)=-\sqrt{2\over -\pi k\tau}e^{-ik\tau}\lf(1-{i\over k\tau} \rt)$, one gets \be
u_k={1\over \sqrt{2k}}e^{-ik\tau}\lf(1-{i\over k\tau} \rt), \ee
which is just the lowest-energy state in de Sitter spacetime (i.e., the BD
vacuum).

Generally, setting the initial state of perturbation
modes as the BD state is essential for inflation to predict the
nearly Gaussian, scale-invariant primordial perturbations. As for
the non-BD states, we will have $|\alpha|>1$ and $|\beta|> 0$,
where $|\beta|^2$ could be interpreted as the number density of
particle excitations \cite{Ford:1997hb}.

The choice of the initial BD state seems natural. However, it is
not the case due to the well-known ``trans-Planckian" problem
\cite{Brandenberger:2000wr,Martin:2000xs}. In the following, based
on the EFT of cosmological perturbations, we will revisit this
problem.

As mentioned, one will get the BD state by requiring the
perturbation modes $u_k$ at $-k\tau\gg 1$ to coincide with the Minkowski
solution. Usually, it is thought that in the infinite past, the
perturbation modes are deep inside the horizon, so that
Eq. (\ref{MS-eq02}) equals the equation of free wavepacket in
flat Minkowski space. However, the so-called Minkowski spacetime
is only ``conformal" Minkowski, which is actually expanding. In
such a spacetime, since the wavelength of perturbation modes
$\lambda \sim a$, which will blueshift back to the past, we will inevitably have $\lambda\lesssim 1/\Lambda$ before some
    $\tau=\tau_{\Lambda}$, where $\Lambda\lesssim M_{\mathrm{P}}$ is the cutoff
scale below which the higher-order derivative operators are
negligible.

When $\lambda\simeq {\cal O}(1/\Lambda)$, the EFT
responsible for (\ref{eq:quadratic}) is no longer robust, and the
neglected higher-order derivative operators, such as (see, e.g., \cite{Cai:2016thi}),
\be L/M_{\mathrm{P}}^2\sim {\lambda_4\over
    \Lambda^2}(R^{(3)})^2+{{\lambda}_6\over
    \Lambda^4}(\nabla_iR^{(3)})^2+\cdots, \label{highorder}\ee will be
no longer negligible, where $R^{(3)}$ is the Ricci scalar on the
3-dimensional spacelike hypersurface and $\lambda_{4,6}\simeq
{\cal O}(1)$ are constants. One is also allowed to take account of the
operators $R^{(3)}_{ij}R^{(3)ij}$ and
$\nabla_iR^{(3)}_{jk}\nabla^iR^{(3)jk}$, which we have omitted for
simplicity. We have put such operators and the higher-order
spatial derivative operators of $R^{(3)}$ into the ellipsis.

Generally,\footnote{One also has considered the operators
    $\sim\nabla_{\mu}\delta K_{\nu\sigma}\nabla^{\mu}\delta K^
    {\nu\sigma}$, which contribute
    $k^6\zeta_k^2$ as well \cite{Ashoorioon:2018uey}.} \be (R^{(3)})^2\sim
k^4\zeta_k^2,\quad {(\nabla_iR^{(3)})^2}\sim k^6\zeta_k^2.\ee The
higher-order derivative operators in (\ref{highorder}) do not
affect the background. Hence, the effective theory of inflation may be
still valid at the background level when the perturbation mode satisfies  $\lambda\lesssim
    1/\Lambda$. However, in this situation, Eq. (\ref{MS-eq02}) will
be modified to \be u_k^{\prime\prime}+\lf(\sim \sum_p {k^{2p}\over
    \Lambda^{2p}}\rt)k^2u_k=0,\label{strong}\ee where $p\geqslant 0$
is some integer. In this case, what the perturbation $u_k$ feels
will be a regime dominated by the higher-order momentum operators
instead of a Minkowski space, since Eq. (\ref{MS-eq02}) is no
longer the equation of free wavepacket in flat Minkowski space.
Therefore, it is not sufficiently reasonable to set the state of
the perturbation modes in the infinite past as the Minkowski
vacuum.


\section{Implications of TCC for pre-inflation }\label{sec:ini}

\subsection{On TCC}

Recently, Bedroya and Vafa proposed the Trans-Planckian Censorship
Conjecture (TCC) \cite{Bedroya:2019snp}, which requires
that the sub-Planckian fluctuations should never cross their
Hubble scale to become classical, otherwise the corresponding EFT
belongs to the swampland
\cite{Obied:2018sgi,Agrawal:2018own,Garg:2018reu,Ooguri:2018wrx}. The TCC
suggests that inflation can only last for a limited $e$-folding
number, i.e., \be \int_{t_i}^{t_f}H dt<\ln{M_{\mathrm{P}}\over
H_{f}}\,,\label{eq:tcc01} \ee where the subscript ``$i$" and
``$f$" represent the beginning and ending of inflation,
respectively. The TCC might be also argued as follows.


Generally, neglecting the higher-order operators in
(\ref{highorder}), we have $|u_k|= 1/\sqrt{2k}$ for the
sub-horizon perturbation modes (noting that it is also
approximately valid for $\lambda\lesssim 1/H$, see also
Sec. \ref{sec:ana}). Thus one will have $|\zeta_k|\sim 1/a$ for such a
perturbation mode, since $|\zeta_k|\sim |u_k|/a$ on sub-horizon scale. It can be found
that at the beginning of inflation, the spectrum of the sub-horizon
perturbation ($\lambda\ll 1/H$) is \be
P_{\zeta}^{1/2}(t_i,k)=P_{\zeta}^{1/2}(t_f,k){a_f\over a_i}\sim
{H_f\over \sqrt{\epsilon_f}M_{\mathrm{P}}}{a_f\over a_i}, \label{tik}\ee
provided this perturbation mode crossed the horizon ($k\simeq a_f
H_f$) at $t_f$. The validity of perturbation theory requires that
$P_{\zeta}^{1/2}(t_i,k)<1$, otherwise the spacetime will be distorted by
lots of black holes on the sub-Hubble scale. According to (\ref{tik}),
we have 
\be \int_{t_i}^{t_f}H dt< \ln{\sqrt{\epsilon_f} M_{\mathrm{P}}\over
H_{f}},\label{Hdt}
\ee 
which, for $\epsilon_f\lesssim 1$, is
consistent with the TCC constraint (\ref{eq:tcc01}).\footnote{It is
interesting that for the tensor perturbation, (\ref{Hdt}) is
exactly same as the TCC constraint (\ref{eq:tcc01}).}

\begin{figure}[tbp]
	\centering 
	\includegraphics[width=.55\textwidth,trim=0 0 0 0,clip]{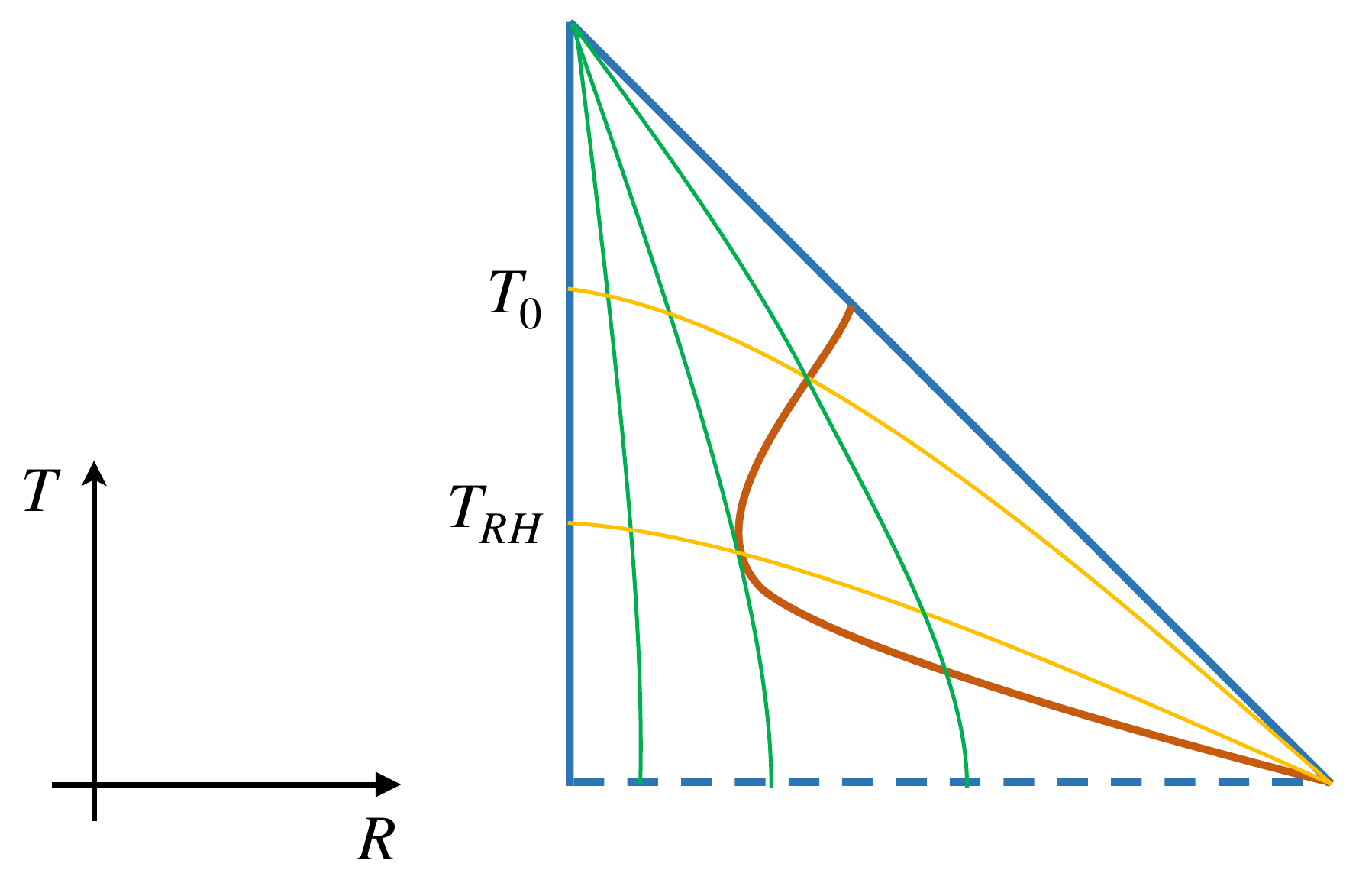}
	\caption{\label{fig-inf} Causal patch of the Friedmann-Lemaitre-Robertson-Walker (FLRW) 
	universe with inflation, where $T$ and $R$ are the timelike and
		radial coordinates, respectively. The dashed line represents the
		initial cosmological singularity, the red solid curve represents
		the comoving horizon (i.e.,
		$1/(aH_{\text{per}})=|z/z^{\prime\prime}|^{1/2}$) for the perturbation
		modes, $T_{RH}$ and $T_0$ represent the reheating surface and
		present, respectively. Initially, all perturbation modes are deep
		inside the horizon. As time goes by, perturbation modes with
		sufficiently large wavelength will exit horizon and eventually
		reenter the horizon. We have assumed that the universe enters into
		the radiation or matter phase after reheating.}
\end{figure}

The TCC might suggest the existence of a pre-inflationary
evolution. The question is whether the pre-inflationary evolution can prepare the required initial condition for observational
(i.e., at the CMB window) inflationary perturbations.
The relevant issue is closely related to the past-incompletion (or
singularity) of inflation \cite{Borde:2001nh}. In inflationary
spacetime, the affine length $\int d\eta_{\zeta}\sim \int c_s^2Q_s
a dt$ of scalar perturbation mode must be finite. As a result, back to the
past, the perturbation modes will inevitably hit the singularity,
as depicted in Fig. \ref{fig-inf}. 
Although it seems acceptable
to set the BD state as the initial state of perturbation modes
below the scale $\Lambda\lesssim M_{\mathrm{P}}$,\footnote{The decay of non-BD state into
BD state through the self-interaction of the perturbations $\sim
\epsilon^2\zeta(\zeta^\prime)^2$ is very inefficient
\cite{ArmendarizPicon:2006pd}, since $\epsilon\ll 1$.
Nevertheless, see also \cite{Jiang:2015hfa,Jiang:2016nok} for further discussions.}
it is not justified either.

\subsection{Minkowski in past-complete pre-inflation }

Recalling the Raychaudhuri equation for the null geodesics, which
is simplified as (assuming the null vector $k^{\mu}$ is orthogonal
to the hypersurfaces, see e.g., \cite{Carroll:2004st}) \be {d
\theta\over d \eta}\leqslant -{\theta^2\over
2}-R_{\mu\nu}k^{\mu}k^{\nu}\,, \label{RE}\ee where $\eta$ is the
affine parameter of the null geodesics and
$\theta=\triangledown_\mu k^{\mu}$ is the expansion parameter.
When the null convergence condition (NCC)
$R_{\mu\nu}k^{\mu}k^{\nu}\geq0$ is satisfied, one has
$d\theta/d\eta+\theta^2/2<0$. In this case,  back along $\eta$ the
null rays will converge, which indicates that the wavelengths of
perturbation modes will be blueshifted, hence $\lambda\lesssim
1/\Lambda$ is inevitable. In Ref. \cite{Borde:1993xh}, the NCC has
been applied to argue the past-incompletion of inflation. Thus,
only if the pre-inflationary era is
past-complete,\footnote{Necessary condition but not sufficient,
see e.g., \cite{Cai:2015nya}.} it is possible to set the initial
state of perturbations in the infinite past (as the Minkowski
vacuum).

In the pre-inflationary era, the past-completion requires that the
converging null rays must be reversed
(i.e., $d\theta/d\eta>0$) at least for some period.
According to (\ref{RE}), during this period the NCC must be
violated, i.e.,
\be R_{\mu\nu}k^{\mu}k^{\nu}<0, \label{NCC}\ee
or equivalently, $\dot{H}>0$, since $R_{\mu\nu}k^{\mu}k^{\nu}=-2{\dot
H}(k^{0})^2$ for the spatially flat FLRW metric.

\begin{figure}[tbp]
    \includegraphics[scale=2,width=0.55\textwidth]{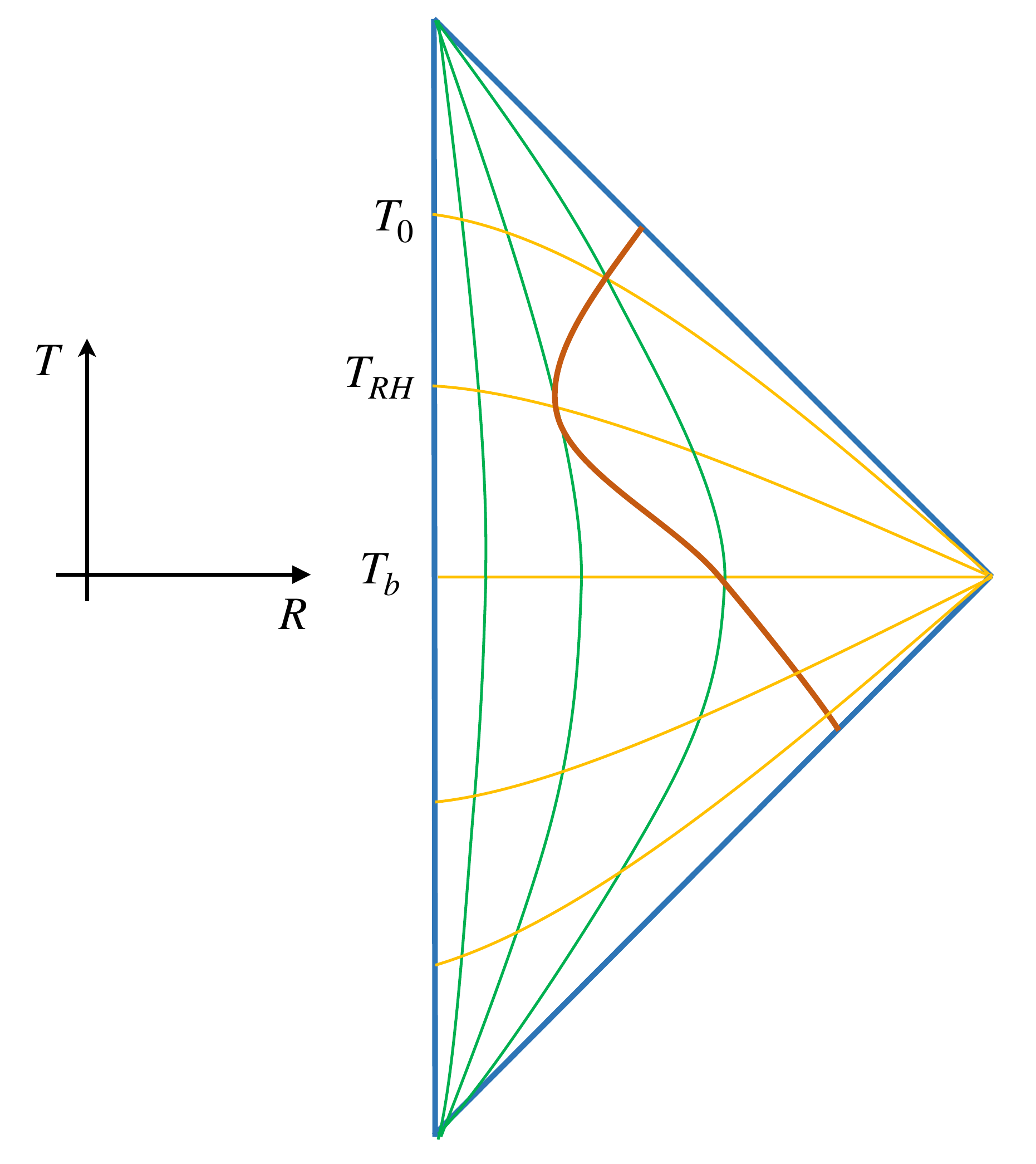}
\caption{Causal patch of the FLRW universe in the bounce
inflation scenario, where $T_b$ represents the time of bounce.}
\label{fig-binf}
\end{figure}

One possibility that satisfies (\ref{NCC}) is the pre-inflationary
bounce, where ${\dot H}>0$ regardless of whether the gravity is GR
or not.
In the corresponding bounce inflation scenario, before the
nonsingular bounce, one may have \be a\sim {1\over
1+\lf({\tau_{c}\over \tau}\rt)^n},
\label{eq:a01}
\ee for $0<n\ll 1$, where
$\tau_c$ is negative. In the infinite past $\tau\rightarrow
-\infty$, $a\sim 1$, we have an asymptotically Minkowski
spacetime. When $|\tau|\ll |\tau_c|$, $a\sim ({\tau/ \tau_c})^n$,
the pre-inflationary universe is slowly contracting (i.e., the ekpyrotic
phase \cite{Khoury:2001wf}) until a nonsingular bounce occurred.
Hereafter it will begin to inflate. The corresponding causal patch
is displayed in Fig. \ref{fig-binf}.

In such a bounce inflation scenario, all perturbation modes
are in a flat Minkowski space in the infinite past. The
wavelengths of the perturbation modes that are interested for
observations satisfy $\lambda\gg 1/\Lambda$, so that the
$k/\Lambda$-suppressed higher-order derivative operators in
(\ref{highorder}) are negligible. In particular, such a
scenario is also consistent with the TCC, since the sub-Planckian
perturbation modes may never have the opportunity to cross their
horizon,\footnote{Strictly, the so-called ``horizon" is actually
``the horizon of the perturbations", which is defined by
$1/H_{\text{per}}=a\sqrt{|z/z^{\prime\prime}|}$. It will be significantly
different from $1/H$, if $Q_s$ is rapidly changing, see also
\cite{Liu:2011ns}.} as long as $a_{f}/a_{b}<M_{\mathrm{P}}/H_{f}$ and $a_{b}$
is sufficiently large (see also \cite{Bedroya:2019tba}),
where the subscript ``$b$" corresponds to the bounce point.

\section{Bunch-Davis at the beginning of inflation}\label{sec-pert}

\subsection{Analytical evaluation}
\label{sec:ana}

In this section, we will focus on the bounce inflation scenario and
evaluate the states of perturbation modes at the beginning of
inflation. The result is also applicable for other scenarios with a
past-complete pre-inflationary era.

Generally, the evolution of the perturbation modes satisfies
Eq. (\ref{MS-eq02}) with $c_s^2$ replaced by ${\cal C}_s^{2}$, where
\be {\cal C}_s^{2}=c_s^2+\lf(\sim \sum_p
{k^{2p}\over \Lambda^{2p}}\rt)\,.\label{eq:cseff} \ee
In the pre-inflationary era, we can safely neglect higher-order
derivative operators and have ${\cal C}_s^{2}=c_s^2$, provided the wavelengths of the perturbation modes
$\lambda\ll 1/\Lambda$ (see also \cite{Brandenberger:2002ty} for related discussions). One can always set $Q_s>0$ and $c_s^2=1$, as in
Refs. \cite{Cai:2016thi,Creminelli:2016zwa,Cai:2017tku,Cai:2017dyi},
so that the ghost or gradient instabilities are avoided.

Considering the pre-inflationary era consists of different phases
with $n=n_j$ $(j\geq 0)$ in Eq. (\ref{eq:a01}), we have \be z_j=\sqrt{2 a^{2} Q_{s}}\sim
(\tau_{R,j}-\tau)^{n_j}\label{z}\ee for the $j$-th phase, where
$n_j\simeq const$. During the bounce, $a$ is almost
constant. However, since $Q_s$ is rapidly changing,
we still could have
Eq. (\ref{z}), see Ref. \cite{Cai:2017pga}. Hence, in the phase $j$,
$z_j^{\prime\prime}/z_j$ can be written as \be \frac{z_j^{\prime
        \prime}}{z_j}={\nu_j^2-{1/4} \over
    {(\tau-\tau_{R,j})^2}}\,,\label{zppbz02} \ee where
$\nu_j=|n_j-1/2|$. Regarding the phases $j$ and $j+1$ as adjacent
phases, we can write down the solutions to Eq. (\ref{MS-eq02}) in
the corresponding phases as \ba u_{k,j}(\tau)&=&{\sqrt{\pi
        (\tau_{R,j}-\tau)}\over
    2}\lf\{\alpha_{j}H_{\nu_{j}}^{(1)}[k(\tau_{R,j}-\tau)]+\beta_{j}H_{\nu_j}^{(2)}[k(\tau_{R,j}-\tau)]
\rt\} ,\,\, (\tau<\tau_{j+1})\,,
\\
u_{k,{j+1}}(\tau)&=&{\sqrt{\pi (\tau_{R,{j+1}}-\tau)}\over
    2}\Big\{\alpha_{j+1}H_{\nu_{j+1}}^{(1)}[k(\tau_{R,j+1}-\tau)]
\nn\\
&\,&\qquad\qquad\qquad\quad
+\beta_{j+1}H_{\nu_{j+1}}^{(2)}[k(\tau_{R,j+1}-\tau)] \Big\}
,\qquad\qquad\qquad (\tau>\tau_{j+1})\,, \label{solution}\ea where
$\alpha_{j(j+1)}$ and $\beta_{j(j+1)}$ are $k$-dependent
coefficients, $\tau_{j+1}$ denotes the conformal time of the juncture of these two phases.

Using the matching conditions
$u_{k,j}(\tau_{j+1})=u_{k,j+1}(\tau_{j+1})$ and
$u_{k,j}'(\tau_{j+1})=u_{k,j+1}'(\tau_{j+1})$ (see e.g., \cite{Joras:2008ck}), we have \ba \left(
\begin{array}{ccc} \alpha_{j+1}\\ \beta_{j+1}
\end{array}\right)
&=& \left(\begin{array}{ccc} {\cal M}_{11}&{\cal
        M}_{12}\\{\cal M}_{21}&{\cal M}_{22}\end{array}\right)
\left(\begin{array}{ccc} \alpha_{j}\\
    \beta_{j}\end{array}\right)\,, \label{Mmetric}\ea with \ba
&\,&{\cal
    M}_{11}={ik\pi\sqrt{\tilde{\tau}_{j+1,1}\tilde{\tau}_{j+1,2}}\over4}\Big[
H^{(1)}_{\nu_j}(k\tilde{\tau}_{j+1,1})H^{(2)}_{\nu_{j+1}-1}(k\tilde{\tau}_{j+1,2})
-H^{(1)}_{\nu_j-1}(k\tilde{\tau}_{j+1,1})H^{(2)}_{\nu_{j+1}}(k\tilde{\tau}_{j+1,2})
\nn\\
&\,&\qquad\qquad\qquad\qquad\qquad\,\,\,\,\, + \lf({2\nu_j-1\over
    2k\tilde{\tau}_{j+1,1}}-{2\nu_{j+1}-1\over 2k\tilde{\tau}_{j+1,2}} \rt)
H^{(1)}_{\nu_j}(k\tilde{\tau}_{j+1,1})H^{(2)}_{\nu_{j+1}}(k\tilde{\tau}_{j+1,2})
\Big]\,,
\\
&\,&{\cal
    M}_{12}={ik\pi\sqrt{\tilde{\tau}_{j+1,1}\tilde{\tau}_{j+1,2}}\over4}\Big[
H^{(2)}_{\nu_j}(k\tilde{\tau}_{j+1,1})H^{(2)}_{\nu_{j+1}-1}(k\tilde{\tau}_{j+1,2})
-H^{(2)}_{\nu_j-1}(k\tilde{\tau}_{j+1,1})H^{(2)}_{\nu_{j+1}}(k\tilde{\tau}_{j+1,2})
\nn\\
&\,&\qquad\qquad\qquad\qquad\qquad\,\,\,\,\, + \lf({2\nu_j-1\over
    2k\tilde{\tau}_{j+1,1}}-{2\nu_{j+1}-1\over 2k\tilde{\tau}_{j+1,2}} \rt)
H^{(2)}_{\nu_j}(k\tilde{\tau}_{j+1,1})H^{(2)}_{\nu_{j+1}}(k\tilde{\tau}_{j+1,2})
\Big]\,,
\\
&\,&{\cal
    M}_{21}={-ik\pi\sqrt{\tilde{\tau}_{j+1,1}\tilde{\tau}_{j+1,2}}\over4}\Big[
H^{(1)}_{\nu_j}(k\tilde{\tau}_{j+1,1})H^{(1)}_{\nu_{j+1}-1}(k\tilde{\tau}_{j+1,2})
-H^{(1)}_{\nu_j-1}(k\tilde{\tau}_{j+1,1})H^{(1)}_{\nu_{j+1}}(k\tilde{\tau}_{j+1,2})
\nn\\
&\,&\qquad\qquad\qquad\qquad\qquad\,\,\,\,\, + \lf({2\nu_j-1\over
    2k\tilde{\tau}_{j+1,1}}-{2\nu_{j+1}-1\over 2k\tilde{\tau}_{j+1,2}} \rt)
H^{(1)}_{\nu_j}(k\tilde{\tau}_{j+1,1})H^{(1)}_{\nu_{j+1}}(k\tilde{\tau}_{j+1,2})
\Big]\,,
\\
&\,&{\cal
    M}_{22}={-ik\pi\sqrt{\tilde{\tau}_{j+1,1}\tilde{\tau}_{j+1,2}}\over4}\Big[
H^{(2)}_{\nu_j}(k\tilde{\tau}_{j+1,1})H^{(1)}_{\nu_{j+1}-1}(k\tilde{\tau}_{j+1,2})
-H^{(2)}_{\nu_j-1}(k\tilde{\tau}_{j+1,1})H^{(1)}_{\nu_{j+1}}(k\tilde{\tau}_{j+1,2})
\nn\\
&\,&\qquad\qquad\qquad\qquad\qquad\,\,\,\,\, + \lf({2\nu_j-1\over
    2k\tilde{\tau}_{j+1,1}}-{2\nu_{j+1}-1\over 2k\tilde{\tau}_{j+1,2}} \rt)
H^{(2)}_{\nu_j}(k\tilde{\tau}_{j+1,1})H^{(1)}_{\nu_{j+1}}(k\tilde{\tau}_{j+1,2})
\Big]\,, \ea where we have defined
$\tilde{\tau}_{j+1,1}=\tau_{R,j}-\tau_{j+1}$ and
$\tilde{\tau}_{j+1,2}=\tau_{R,j+1}-\tau_{j+1}$.

When $k\tilde{\tau}_{j+1,1}$ and $k\tilde{\tau}_{j+1,2}\gg1$, we have \ba
&\,&{\cal M}_{11}\approx \lf(1-ix \rt)e^{i\theta_{j+1} }\,,
\,{\cal M}_{12}\approx xe^{-i{\tilde\theta}_{j+1} } \,,
\,{\cal M}_{21}\approx xe^{i{\tilde\theta}_{j+1}}\,,
\,{\cal M}_{22}\approx \lf(1+ix\rt) e^{-i\theta_{j+1}}, \ea 
where
$x={1-2\nu_{j} \over 4k\tilde{\tau}_{j+1,1}}-{1-2\nu_{j+1} \over
        4k\tilde{\tau}_{j+1,2}}$,
$\theta_{j+1}=k(\tilde{\tau}_{j+1,1}-\tilde{\tau}_{j+1,2}){-\frac{\pi }{2}
    (\nu_{j}-\nu_{j+1})}$, ${\tilde
    \theta_{j+1}}=k(\tilde{\tau}_{j+1,1}+\tilde{\tau}_{j+1,2}){-\frac{ \pi }{2}
    (\nu_{j}+\nu_{j+1})}$. Thus, considering the perturbation modes
what we concern satisfy $k\tilde{\tau}_{j+1,1}$ and $k\tilde{\tau}_{j+1,2}\gg1$,
we have \ba \left(
\begin{array}{ccc} \alpha_{j+1}\\ \beta_{j+1}
\end{array}\right)
&\approx& \left(
\begin{array}{ccc} e^{i\theta_{j+1}}\lf[1+{\cal O}\lf({1\over k\tilde{\tau}}
    \rt)\rt] & {\cal O}\lf({1\over k\tilde{\tau}} \rt)
    \\
    {\cal O}\lf({1\over k\tilde{\tau}} \rt) & e^{-i\theta_{j+1}} \lf[1+{\cal
        O}\lf({1\over k\tilde{\tau}} \rt)\rt]
\end{array}\right)
\left(\begin{array}{ccc} \alpha_{j}\\
    \beta_{j}\end{array}\right)\,, \label{iii1} \ea where $\tilde{\tau}$
correspond to $\tilde{\tau}_{j+1,1}$ or $\tilde{\tau}_{j+1,2}$ and $j\geq0$.

Define the past-infinite phase and the inflation as the phases
$j=0$ and $j=i$, respectively, with Eq. (\ref{iii1}), we
straightforwardly find that \ba \left(
\begin{array}{ccc} \alpha_{i}\\ \beta_{i}
\end{array}\right)
&\approx& \left(
\begin{array}{ccc}
    e^{i\sum_{j=1}^{i} \theta_j}\lf[1+{\cal O}\lf({1\over k\tilde{\tau}}
    \rt)\rt] & {\cal O}\lf({1\over k\tilde{\tau}} \rt)
    \\
    {\cal O}\lf({1\over k\tilde{\tau}} \rt) & e^{-i\sum_{j=1}^{i}
        \theta_j}\lf[1+{\cal O}\lf({1\over k\tilde{\tau}} \rt)\rt]
\end{array}\right)
\left(\begin{array}{ccc} \alpha_{0}\\
    \beta_{0}\end{array}\right)\,, \label{iii2} \ea where \be
\sum_{j=1}^{i}
\theta_j=-{\pi\over2}(\nu_0-\nu_{i})+k\sum_{j=1}^{i}(\tilde{\tau}_{j,1}-\tilde{\tau}_{j,2}),\ee
for the perturbation modes satisfying $k\tilde{\tau}\gg 1$ (i.e.,
$k\tilde{\tau}_{j,1}\gg 1$ and $k\tilde{\tau}_{j,2}\gg 1$ for $j=1,2,\cdots, i$),
which actually equals requiring that the corresponding
perturbation modes have not ever exited its horizon $1/H_{\text{per}}$ in the
pre-inflationary era. Therefore, for the perturbation modes with
$\lambda\ll 1/H_{i}$ at the beginning of inflation,  $\alpha_{i}(k)$ and $\beta_{i}(k)$ acquire only some phase
shift, compared
with their initial values $\alpha_{0}(k)$ and $\beta_{0}(k)$ in the infinite
past.

In the past infinity, if the universe is flat and Minkowskian,
naturally the perturbation mode is in its Minkowski vacuum, i.e.,
$|\alpha_{0}|=1$ and $\beta_{0}=0$.  Hence, \be |\alpha_{i}|\approx
1+{\cal O}\lf({1\over k\tilde{\tau}} \rt)\,,\qquad |\beta_{i}|\approx
{\cal O}\lf({1\over k\tilde{\tau}} \rt) \label{eq:ab} \ee for $k\tilde{\tau}\gg1$. Therefore, at
the beginning of inflation, the perturbation modes with the
wavelengths $\lambda\ll 1/H_{i}$ (the Hubble scale of inflation)
behave themselves as they are in the BD state, which will be
responsible for the nearly scale-invariant density perturbations. As
for the modes with $\lambda\gtrsim 1/H_{i}$, they are in the non-BD
state at the beginning of inflation, hence possibly explaining the
power deficit of the CMB TT-spectrum at low multipoles (e.g., \cite{Qiu:2015nha,Ni:2017jxw}) and the dip
at multipole $l\sim 20$ (see e.g., \cite{Cai:2017pga} and
references therein). A similar analysis can be carried out for the primordial tensor perturbations.

It should be pointed out that the TCC's prediction of an extremely small tensor-to-scalar ratio (see e.g., \cite{Bedroya:2019tba,Mizuno:2019bxy}) implicitly requires that the perturbation modes are in the BD states at the beginning of inflation.
Therefore, in our pre-inflation model, 
the tensor-to-scalar ratio should also be bounded similarly due to the upper bound on the inflationary energy scale put by the TCC, provided the power spectrum of tensor perturbations takes the standard form.

\subsection{Numerical simulation}

Defining
\be C(k)={k\over  2}\lf( u_ku_k^*+{1\over
k^2}u_k'u_k^{*\prime}\rt)+{1\over 2}\,,\qquad (\tau\approx \tau_{i})\,,
\ee
we have \be
C(k)=|\alpha_{i}|^2+{\cal O}\lf({1\over(-k/{\cal
H}_{i})^2}\rt)\simeq |\alpha_{i}|^2\, \ee
for $\nu_{i}\approx 3/2$ and $-k\tau_{i}\gg 1$, where ${\cal
H}_{i}=aH_{i}=\tau_{i}$ and the Wronskian condition has been used. We will use $|C(k)|=1$ as the criterion to
check whether the perturbation modes with $\lambda\ll 1/H_{i}$
at the beginning of inflation are in the BD state.

We consider the bounce inflation model proposed in
Ref. \cite{Cai:2017pga} (see Appendix \ref{Sec:Ap}), where around the bounce we
have ${\cal C}_s^2=c_s^2=1$ and \be Q_s=A_Q\lf[B-\tanh\lf( {t\over
    t_*}\rt) \rt]\,\label{Qs} \ee with constant $A_Q$ and $B$. As
mentioned, in Eq. (\ref{MS-eq02}), $z=\sqrt{2a^2Q_s}$ depends not
only on $a$ but also on $Q_s$. During the ekpyrotic contraction,
$z^{\prime\prime}/z\approx 0/\tau^2$, since both $a$ and $Q_s$
are nearly constant. During the inflation,
$z^{\prime\prime}/z\approx 2/\tau^2$, which is well-known. Around
$t_*$, $Q_s$ has a rapid change, but $H\simeq H_b^2(t-t_b)$ around
the bounce point $t_b$, hence \be a\sim e^{H_b^2(t-t_b)^2/2}\simeq
1+H_b^2(t-t_b)^2/2\ee is nearly constant for $t-t_b\ll 1/H_b$.
As a result, we have $z^{\prime\prime}/z\approx -(\tau-\tau_R)^{-2}/4$  with $\tau_R$ being some reference time.
Therefore, all phases are actually consistent with
(\ref{zppbz02}). As expected, small deviation will not make any
qualitative difference for our result.

With the background of bounce inflation in Ref. \cite{Cai:2017pga}
(see Fig. \ref{fig:H} in Appendix \ref{Sec:Ap}), we plot
$|C(k)|$ in Fig. \ref{fig:Ck} while $A_Q=3$ and $B=2$. We can see that for sufficiently large
$k$ (i.e., $k^2\gg z^{\prime\prime}/z\approx a^2H^2_{i}$), we will have $|C(k)|= 1$.
Thus, the perturbation modes with
wavelengths $\lambda\ll 1/H_{i}$ at the beginning of inflation will be naturally in the BD state.

As for the perturbation modes
with wavelengths comparable or even larger than the Hubble horizon
(i.e., $\lambda \gtrsim 1/H_i$) at the beginning of inflation,
they would be in the non-BD state. These modes exit horizon
during the pre-inflationary era and carry the distinctive
pre-inflationary signatures that can be tested by the CMB
observations. It should be pointed out that if inflation lasted
more than 65 $e$-folds, it would become a challenge to keep the
observational signatures of primordial perturbations generated in the
pre-inflationary era. However, due to the bound of TCC, inflation of less
$e$-folds is favored. As a result, the large scale CMB window will be sensitive to testing the pre-inflationary evolution.
Theoretically, the particular range of the scale of such a window
is model-dependent, as can be inferred from Eq. (\ref{eq:ab}).

\begin{figure}[tbp]
    \includegraphics[scale=2,width=0.55\textwidth]{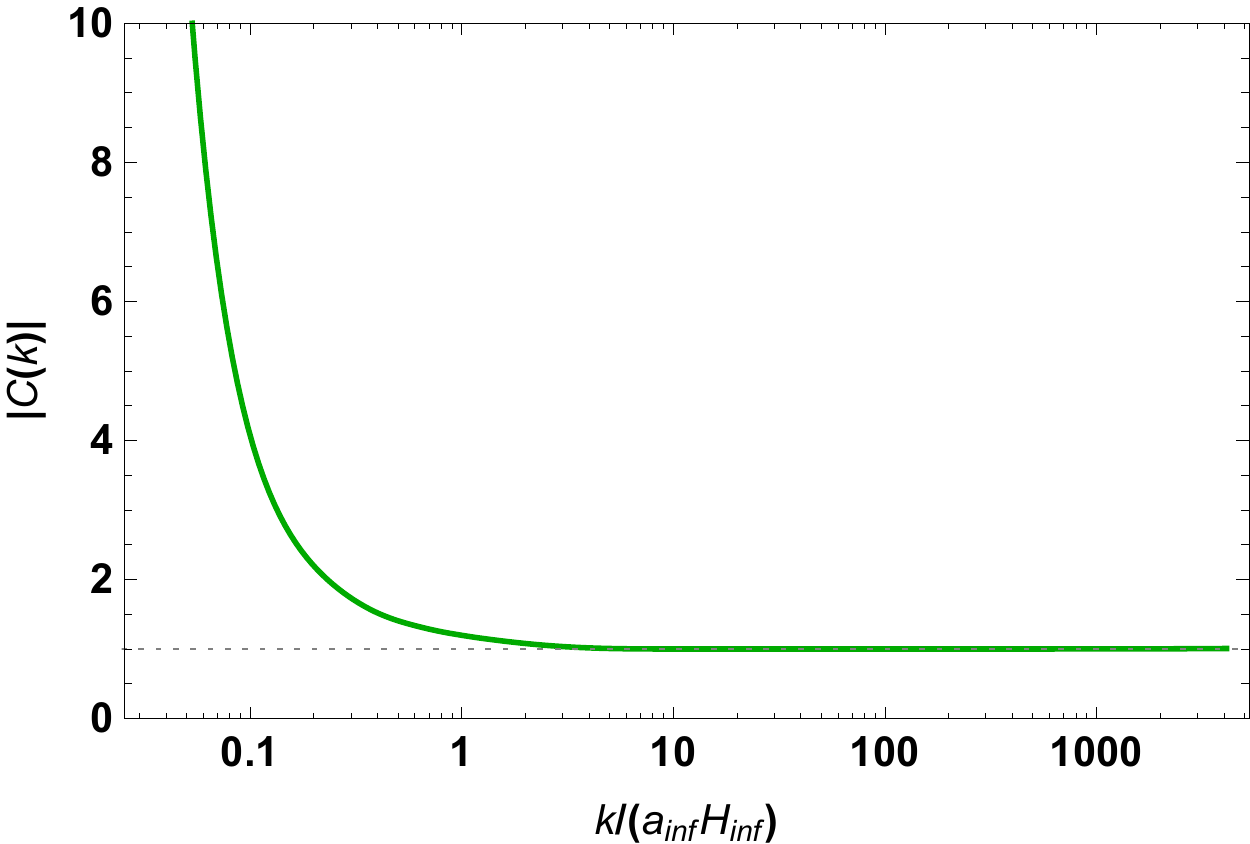}
    \caption{$|C(k)|\rightarrow1$ for $k\gg a_\text{inf}H_\text{inf}$. Thus, the
        perturbation modes with wavelengths much shorter than the Hubble
        scale at the beginning of inflation behave as they are in the BD
        vacuum. } \label{fig:Ck}
\end{figure}

\subsection{Trans-Planckian effect}

So far we have neglected the higher-order momentum corrections in
(\ref{eq:cseff}). However, it is interesting to see what will
happen if they are not negligible.

Generally, with (\ref{eq:cseff}), the solution to
Eq. (\ref{MS-eq02}) can be given as \ba u_{k,j}(\tau)&=&{\sqrt{\pi
        (\tau_{R,j}-\tau)}\over 2}\Big\{\alpha_{j}H_{\nu_{j}}^{(1)}[{\cal
    C}_{s,j} k(\tau_{R,j}-\tau)]
\nn\\
&\,&\qquad\qquad\qquad +\beta_{j}H_{\nu_j}^{(2)}[{\cal C}_{s,j}
k(\tau_{R,j}-\tau)] \Big\} ,\,\quad (\tau<\tau_{j+1})\,.\ea
By matching phase $j$ with phase $j+1$, we can
get (\ref{Mmetric}) with \ba
&\,&{\cal
    M}_{11}={ik\pi\sqrt{\tilde{\tau}_{j+1,1}\tilde{\tau}_{j+1,2}}\over4}\Big[
\lf({2\nu_j-1\over 2k\tilde{\tau}_{j+1,1}}-{2\nu_{j+1}-1\over
    2k\tilde{\tau}_{j+1,2}} \rt) H^{(1)}_{\nu_j}({\cal C}_{s,j}
k\tilde{\tau}_{j+1,1})H^{(2)}_{\nu_{j+1}}({\cal C}_{s,j+1}k\tilde{\tau}_{j+1,2})
\nn\\
&\,&\qquad\qquad\qquad\qquad\qquad\,\,\,\,\, +{\cal C}_{s,j+1}
H^{(1)}_{\nu_j}({\cal C}_{s,j}
k\tilde{\tau}_{j+1,1})H^{(2)}_{\nu_{j+1}-1}({\cal C}_{s,j+1}
k\tilde{\tau}_{j+1,2})
\nn\\
&\,&\qquad\qquad\qquad\qquad\qquad\,\,\,\,\, -{\cal
    C}_{s,j}H^{(1)}_{\nu_j-1}({\cal
    C}_{s,j}k\tilde{\tau}_{j+1,1})H^{(2)}_{\nu_{j+1}}({\cal
    C}_{s,j+1}k\tilde{\tau}_{j+1,2}) \Big]\,,
\\
&\,&{\cal
    M}_{12}={ik\pi\sqrt{\tilde{\tau}_{j+1,1}\tilde{\tau}_{j+1,2}}\over4}\Big[
\lf({2\nu_j-1\over 2k\tilde{\tau}_{j+1,1}}-{2\nu_{j+1}-1\over
    2k\tilde{\tau}_{j+1,2}} \rt) H^{(2)}_{\nu_j}({\cal C}_{s,j}
k\tilde{\tau}_{j+1,1})H^{(2)}_{\nu_{j+1}}({\cal C}_{s,j+1} k\tilde{\tau}_{j+1,2})
\nn\\
&\,&\qquad\qquad\qquad\qquad\qquad\,\,\,\,\, +{\cal C}_{s,j+1}
H^{(2)}_{\nu_j}({\cal C}_{s,j}
k\tilde{\tau}_{j+1,1})H^{(2)}_{\nu_{j+1}-1}({\cal C}_{s,j+1}
k\tilde{\tau}_{j+1,2})
\nn\\
&\,&\qquad\qquad\qquad\qquad\qquad\,\,\,\,\, -{\cal
    C}_{s,j}H^{(2)}_{\nu_j-1}({\cal C}_{s,j}
k\tilde{\tau}_{j+1,1})H^{(2)}_{\nu_{j+1}}({\cal C}_{s,j+1} k\tilde{\tau}_{j+1,2})
\Big]\,,
\\
&\,&{\cal
    M}_{21}={-ik\pi\sqrt{\tilde{\tau}_{j+1,1}\tilde{\tau}_{j+1,2}}\over4}\Big[
\lf({2\nu_j-1\over 2k\tilde{\tau}_{j+1,1}}-{2\nu_{j+1}-1\over
    2k\tilde{\tau}_{j+1,2}} \rt) H^{(1)}_{\nu_j}({\cal C}_{s,j}
k\tilde{\tau}_{j+1,1})H^{(1)}_{\nu_{j+1}}({\cal C}_{s,j+1} k\tilde{\tau}_{j+1,2})
\nn\\
&\,&\qquad\qquad\qquad\qquad\qquad\,\,\,\,\, +{\cal C}_{s,j+1}
H^{(1)}_{\nu_j}({\cal C}_{s,j}
k\tilde{\tau}_{j+1,1})H^{(1)}_{\nu_{j+1}-1}({\cal C}_{s,j+1}
k\tilde{\tau}_{j+1,2})
\nn\\
&\,&\qquad\qquad\qquad\qquad\qquad\,\,\,\,\, -{\cal
    C}_{s,j}H^{(1)}_{\nu_j-1}({\cal C}_{s,j}
k\tilde{\tau}_{j+1,1})H^{(1)}_{\nu_{j+1}}({\cal C}_{s,j+1} k\tilde{\tau}_{j+1,2})
\Big]\,,
\\
&\,&{\cal
    M}_{22}={-ik\pi\sqrt{\tilde{\tau}_{j+1,1}\tilde{\tau}_{j+1,2}}\over4}\Big[
\lf({2\nu_j-1\over 2k\tilde{\tau}_{j+1,1}}-{2\nu_{j+1}-1\over
    2k\tilde{\tau}_{j+1,2}} \rt) H^{(2)}_{\nu_j}({\cal C}_{s,j}
k\tilde{\tau}_{j+1,1})H^{(1)}_{\nu_{j+1}}({\cal C}_{s,j+1} k\tilde{\tau}_{j+1,2})
\nn\\
&\,&\qquad\qquad\qquad\qquad\qquad\,\,\,\,\, +{\cal C}_{s,j+1}
H^{(2)}_{\nu_j}({\cal C}_{s,j}
k\tilde{\tau}_{j+1,1})H^{(1)}_{\nu_{j+1}-1}({\cal C}_{s,j+1}
k\tilde{\tau}_{j+1,2})
\nn\\
&\,&\qquad\qquad\qquad\qquad\qquad\,\,\,\,\, -{\cal
    C}_{s,j}H^{(2)}_{\nu_j-1}({\cal C}_{s,j}
k\tilde{\tau}_{j+1,1})H^{(1)}_{\nu_{j+1}}({\cal C}_{s,j+1} k\tilde{\tau}_{j+1,2})
\Big]\,.  \ea Thus, for the perturbation modes satisfying
$k\tilde{\tau}_{j+1,1}$ and $k\tilde{\tau}_{j+1,2}\gg1$, we have \ba \left(
\begin{array}{ccc} \alpha_{j+1}\\ \beta_{j+1}
\end{array}\right)
&\approx& \left(
\begin{array}{ccc} e^{i\theta_{j+1}}\lf[
    {  {\cal C}_{s,j}+{\cal C}_{s,j+1} +{\cal O}\lf({1\over k\tilde{\tau}} \rt) \over 2\sqrt{ {\cal C}_{s,j}{\cal C}_{s,j+1} } }\rt]
    &
    -e^{-i{\tilde\theta}_{j+1}}\lf[
    {  i({\cal C}_{s,j}-{\cal C}_{s,j+1}) +{\cal O}\lf({1\over k\tilde{\tau}} \rt) \over 2\sqrt{ {\cal C}_{s,j}{\cal C}_{s,j+1} } }\rt]
    \\
    e^{i{\tilde\theta}_{j+1}}\lf[
    {  i({\cal C}_{s,j}-{\cal C}_{s,j+1}) +{\cal O}\lf({1\over k\tilde{\tau}} \rt) \over 2\sqrt{ {\cal C}_{s,j}{\cal C}_{s,j+1} } }\rt]
    &
    e^{-i{\theta}_{j+1}}\lf[
    {  {\cal C}_{s,j}+{\cal C}_{s,j+1} +{\cal O}\lf({1\over k\tilde{\tau}} \rt) \over 2\sqrt{ {\cal C}_{s,j}{\cal C}_{s,j+1} } }\rt]
\end{array}\right)
\left(\begin{array}{ccc} \alpha_{j}\\
    \beta_{j}\end{array}\right), \label{iii15} \ea which reduces to
(\ref{iii1}) when ${\cal C}_s^2=1$, where $\theta_{j+1}=k({\cal
    C}_{s,j} \tilde{\tau}_{j+1,1}-{\cal C}_{s,j+1} \tilde{\tau}_{j+1,2}){-\frac{ \pi
    }{2} (\nu_{j}-\nu_{j+1})}$ and ${\tilde \theta}_{j+1}=k({\cal
    C}_{s,j} \tilde{\tau}_{j+1,1}+{\cal C}_{s,j+1} \tilde{\tau}_{j+1,2}){-\frac{ \pi
    }{2} (\nu_{j}+\nu_{j+1})}$.

Consider such a pre-inflationary era, in which ${\cal C}_s^2=(
k/\Lambda)^{2p}\gg 1$ for the phase $j$, while ${\cal C}_s^2=1$ in
other phases. According to (\ref{iii15}), we straightforwardly have \ba
\left(
\begin{array}{ccc} \alpha_{j+1}\\ \beta_{j+1}
\end{array}\right)
&\approx& {k^p\over 2\Lambda^p}\left(
\begin{array}{ccc} e^{i\theta_{j+1}}
    &
    -ie^{-i{\tilde\theta}_{j+1}}
    \\
    ie^{i{\tilde\theta}_{j+1}}
    &
    e^{-i{\theta}_{j+1}}
\end{array}\right)
\left(\begin{array}{ccc} \alpha_{j}\\
    \beta_{j}\end{array}\right)\,.\ea
This suggests that even if we have $|\alpha_{0}|=1$ and
$\beta_{0}=0$ initially or in the infinite past,  both
$\alpha_{i}$ and $\beta_{i}$ will be altered significantly as \be
|\alpha_{i}|\sim {k^p\over 2\Lambda^p}\,,\qquad |\beta_{i}|\sim
{k^p\over 2\Lambda^p} \ee for $k\tilde{\tau}\gg1$, after passing
through one Planckian regime ($k\gg \Lambda$). Thus, in this case,
the initial state of perturbation modes on sub-horizon scale will
be no longer (actually far away from) the BD state at the
beginning of inflation. This is actually equivalent to the
``trans-Planckian problem" of inflation. The initial state of
perturbation modes at the beginning of inflation being the
BD state requires that the wavelengths of the perturbation modes
(at the CMB window) must satisfy $\lambda\gg 1/\Lambda$ at any
moment in the pre-inflationary era, so that the higher-order
spatial derivative operators may be neglected. This is
consistent with the condition of TCC. In certain sense, such a
pre-inflationary era should be past-complete, otherwise the phase
with $\lambda< 1/\Lambda$ is inevitable.

\section{Conclusion}\label{conclusion}

Due to the past-incompleteness of inflation, the pre-inflationary
physics must be considered to set the initial states for the
inflationary perturbations.
We discussed the implication of TCC for the initial states
of the primordial perturbations and pointed out that it is possible to
consistently set the initial state of perturbation modes in the
infinite past, only if a past-complete pre-inflationary era
exists.

In the past-complete pre-inflationary era, the spacetime may be Minkowskian in the infinite past. Hence, it is natural to
pick out the Minkowski vacuum state as the initial state of
perturbations. Moreover, in such a scenario, the
sub-Planckian perturbation modes can never cross their horizon,
which is also consistent with the TCC. We calculate the evolution
of the perturbation modes in such a pre-inflationary era, and show
that the perturbation modes will behave as they are in the BD
state at the beginning of inflation, as long as their wavelengths
$\lambda\ll 1/H$ at that time. Therefore, a past-complete
pre-inflationary evolution may automatically prepare the initial
state required for the inflationary perturbations.

The past-complete pre-inflationary era might have left the imprint
on CMB. One possibility of such scenarios is the bounce inflation
\cite{Piao:2003zm,Liu:2013kea,Qiu:2015nha}. As showed recently in
\cite{Cai:2017pga}, the pre-inflationary bounce might be
responsible for the power deficit of the CMB TT-spectrum at low
multipoles and the dip at $l\sim 20$. Relevant issues are
interesting for further study. In addition, it is also possible that the
universe, as well as the initial state of perturbation modes, was
created quantumly \cite{Vilenkin:1982de,Hartle:1983ai}. In such a
scenario, how the perturbation is in its lowest-energy state after the creation is still under study, see recent
Refs. \cite{Feldbrugge:2017fcc,DiazDorronsoro:2017hti,Vilenkin:2018dch,Bojowald:2018gdt,DiTucci:2019dji}.

\acknowledgments

Y. C. is funded by the China Postdoctoral Science Foundation (Grant No. 2019M650810) and NSFC (Grant No. 11905224). Y. S. P. is supported by NSFC (Grant Nos. 11575188 and 11690021).
This project is supported by the University of Chinese Academy of Sciences.

\appendix

\section{Bounce inflation model}
\label{Sec:Ap}

The stable nonsingular bounce inflation model has been implemented
in the beyond Horndeski theory, as inspired by the EFT of
nonsingular cosmologies
\cite{Cai:2016thi,Creminelli:2016zwa,Cai:2017tku,Cai:2017dyi,Kolevatov:2017voe}. In
this Appendix, we briefly review it, see \cite{Cai:2017pga} for
the details.

Our effective Lagrangian for the bounce inflation is
\ba\label{action01} L\sim & & \underbrace{{M_{\mathrm{P}}^2\over
        2}R-{M_{\mathrm{P}}^2\over 2}X-V(\phi)} \nn\\ & & \text{\textit{Contraction}}+\text{\textit{Inflation}}\nn\\
&+& \quad\quad\underbrace{{\tilde
        P}(\phi,X)}\quad\quad\quad\quad\quad\quad +\underbrace{L_{\delta
        g^{00} R^{(3)}}} \quad\quad+L_{\delta K \delta g^{00} }\,, \\
& & \text{\textit{(Ghost free) Bounce}} \quad\quad\,\,
\text{\textit{Removing}} ~ c_s^2<0 \nn\ea where
\ba  L_{\delta g^{00} R^{(3)}}& =& {f_1(\phi)\over 2}\delta g^{00} R^{(3)}\nn\\
&=& {f\over 2}R -{X\over 2} \int f_{\phi\phi}d \ln X
-\lf(f_\phi+\int {f_\phi \over 2}d\ln X\rt)\Box\phi \nn\\
& \, & + {f\over 2X}\lf[\phi_{\mu\nu}\phi^{\mu\nu}-(\Box
\phi)^2\rt]-{f-2Xf_X\over
    X^2}\lf[\phi^\mu\phi_{\mu\rho}\phi^{\rho\nu}\phi_\nu-(\Box
\phi)\phi^\mu\phi_{\mu\nu}\phi^\nu\rt] \,,\label{g00R3-01} \\ L_{\delta K \delta
    g^{00} }&=& {g_1(\phi)\over2}\delta K\delta g^{00}
\nn\\
&=&{g\over
    2}{1\over\sqrt{-X}}\lf({\phi^\mu\phi_{\mu\nu}\phi^\nu\over X}-\Box
\phi \rt) -{3\over2}g {\cdot  h}\,,\label{Kg00-01}\ea
$R^{(3)}\delta g^{00}$ and $\delta K\delta g^{00}$ are the EFT
operators which will modify $c_s^2$ and $Q_s$ in
Eq. (\ref{MS-eq02}), $\phi_\mu=\nabla_\mu\phi$,
$\phi^\mu=\nabla^\mu\phi$,
$\phi_{\mu\nu}=\nabla_\nu\nabla_\mu\phi$, $X=\phi_{\mu}\phi^\mu$
and $\Box\phi=\phi^\mu_{~\mu}$. Here, the coefficients $f$ and $g$
depend on $\phi$ and $X$, while $h$ and $f_2$ depend only on
$\phi$. In certain conditions (see \cite{Cai:2017pga} for
details), $f$, $g$ and $h$ can be set to be vanishing at the
background level.

\begin{figure}[tbp]
    \includegraphics[scale=2,width=0.60\textwidth]{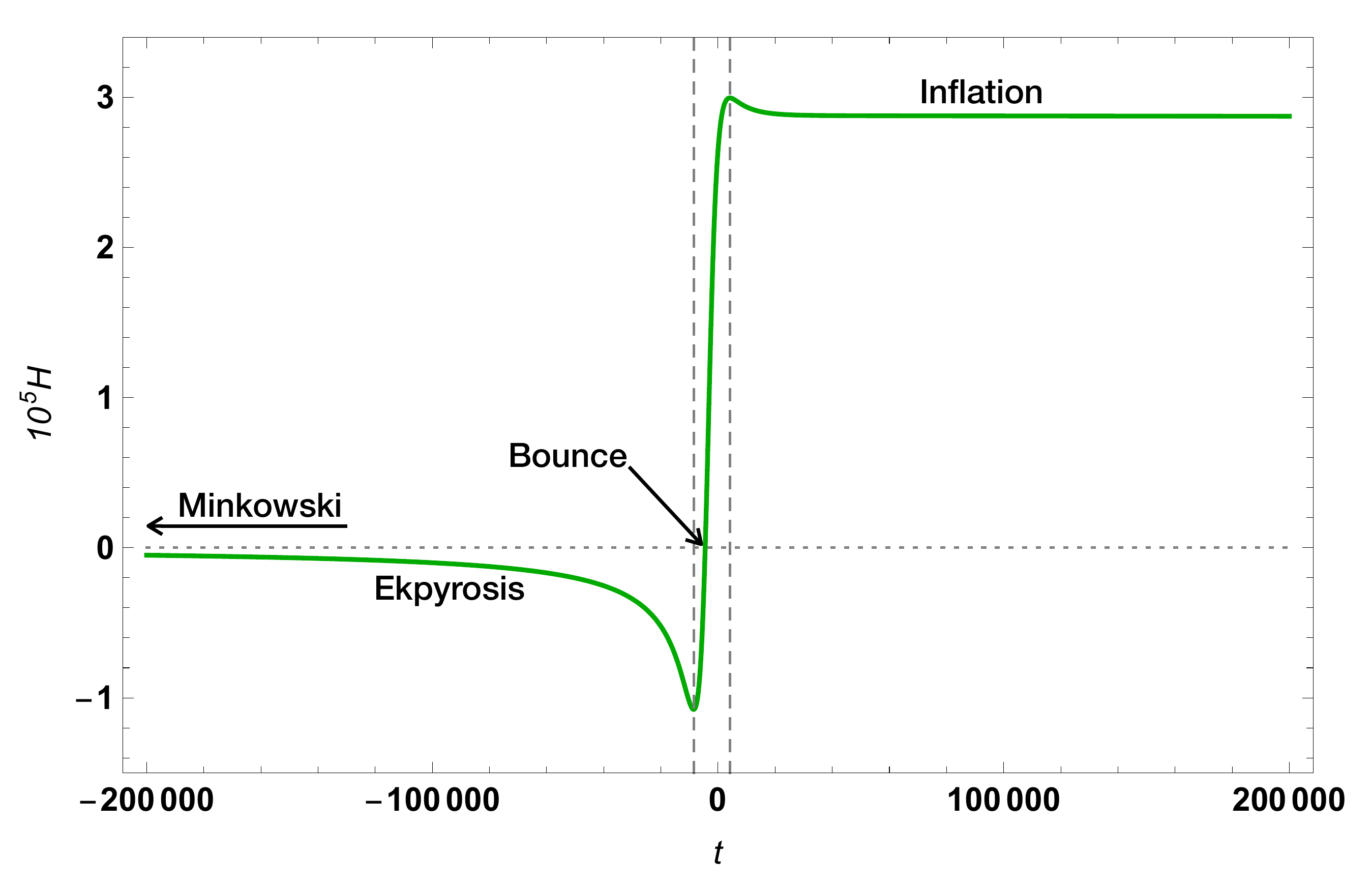}
\caption{The evolution of the Hubble parameter $H$ in bounce
inflation model, where we have set $M_{\mathrm{P}}=1$. }\label{fig:H}
\end{figure}

As shown in Ref. \cite{Cai:2017pga}, with the operators
$R^{(3)}\delta g^{00}$ and $\delta K\delta g^{00}$, we can have
$c_s^2=1$ and $Q_s$ satisfying Eq. (\ref{Qs}). In
\cite{Cai:2017pga}, the background solution to bounce inflation
has been displayed. Here, we only show the evolution of the background
(the Hubble parameter $H$) in Fig. \ref{fig:H}, which will be used
in Sec. \ref{sec-pert}.

\end{document}